\documentclass[a4paper,fleqn]{cas-dc}

\usepackage[numbers]{natbib}
\usepackage{upgreek}
\usepackage{pdfpages}
\usepackage{float}
\usepackage[normalem]{ulem}

\def\tsc#1{\csdef{#1}{\textsc{\lowercase{#1}}\xspace}}
\tsc{WGM}
\tsc{QE}

\begin{document}
\let\WriteBookmarks\relax
\def\floatpagepagefraction{1}
\def\textpagefraction{.001}

\shorttitle{}    

\shortauthors{D. Kolenaty et~al.}  

\title [mode = title]{Corrosion-resistant and conductive Ti--Nb--O coatings tailored for ultra-low Pt-loaded BPPs and PTLs in PEM electrolyzers}  



\author[]{David Kolenatý}[type=editor,
                        orcid=0000-0003-3651-2473]
\cormark[1]
\ead{kolenaty@kfy.zcu.cz}

\author{Jiří Čapek}[orcid=0000-0002-3267-7708]

\author{Stanislav Haviar}[orcid=0000-0001-6926-8927]
\author{Jiří Rezek}[orcid=0000-0002-2698-8753]
\author{Radomír Čerstvý}[orcid=0000-0001-8507-6642]
\author{Akash Kumar}[orcid=0009-0003-8302-1105]
\author{Kalyani Shaji}[orcid=0009-0008-9320-477X]
\author{Mariia Zhadko}[orcid=0000-0002-7979-9550]
\author{Petr Zeman}[orcid=0000-0001-8742-4487]

\affiliation[]{
    organization={Department of Physics and NTIS -- European Centre of Excellence, University of West Bohemia in Pilsen},
    addressline={Univerzitní 8}, 
    city={Pilsen},
    postcode={301 00}, 
    country={Czech Republic}}
            
\begin{abstract}
We develop highly corrosion-resistant and conductive Ti--Nb--O coatings for metallic components---bipolar plates (BPPs) and porous transport layers (PTLs)---in PEM water electrolyzers. Using reactive high-power impulse magnetron sputtering (HiPIMS), we deposit compact 200~nm bilayer coatings onto SS316L substrates, systematically tailoring their composition. By precisely controlling oxygen partial pressure and Nb/Ti ratio, we adjust stoichiometry and structure, directly affecting electrical resistivity and corrosion resistance. We examine interfacial contact resistance (ICR) and electrochemical parameters before and after accelerated corrosion testing. Optimized coatings exhibit resistivity on the order of $10^{-4}~\Omega\text{·}\mathrm{cm}$ and extremely low corrosion current densities ($J_\mathrm{corr} = 0.01$–$0.08~\upmu\mathrm{A}/\mathrm{cm}^2$), well below the U.S. DOE 2026 target. Most importantly, these coatings enable the ICR target after accelerated corrosion testing with a Pt overlayer as thin as 5~nm, reducing Pt loading by up to two orders of magnitude compared to conventional approaches.
\end{abstract}


\begin{highlights}
    \item 200~nm Ti--Nb--O bilayer coatings for BPPs and PTLs made by HiPIMS.
    \item Tailored composition and structure yield compact, conductive coatings.
    \item DOE ICR target met after accelerated corrosion test with 5~nm Pt overlayer.
    \item Pt loading reduced by up to two orders of magnitude versus conventional coatings.
    \item Coatings enable durable PEM electrolyzers with less reliance on precious metals.
\end{highlights}



\date{}
\maketitle

\section{Introduction}\label{sec:level1}
\par With the growing global energy consumption, the human influence on the greenhouse effect and resulting climate change becomes increasingly evident. This stresses the urgent need for the “decarbonization” of the global energy system \cite{Ritchie2023}. Green hydrogen appears to play an essential role in the transition from limited fossil fuels to unlimited renewable energy sources \cite{Dunn2002,Chu2012}. 

\par The need for flexible energy storage solutions has significantly driven the development of Proton Exchange Membrane (PEM) electrolysis in recent years \cite{Gotz2016,Buttler2018}. The main advantages of PEM electrolysis compared to conventional alkaline electrolysis at similar efficiency include high current density allowing compact design and high-pressure operation, shorter start-up time, no need for stand-by protective current (full load range) \cite{Buttler2018,Ayers2010}, and very high purity of produced hydrogen \cite{Buttler2018,Carmo2013} providing better coupling with dynamic and intermittent systems. However, investment in this technology is currently more expensive than in alkaline systems owing to the costs for the membrane and the high loading of precious metals. Therefore, technological innovations such as an increase in current density and a reduction of expensive materials are necessary to achieve competitive capital costs \cite{Buttler2018,Carmo2013}.

\par The porous transport layers (PTLs) and bipolar plates (BPPs) are multifunctional metallic components of PEM electrolyzer stacks, accounting for the main fraction of the system’s total weight and cost. These components facilitate the transport of water, electrons, gases, and heat within the electrolysis cells. BPPs also separate individual cells, transport electrons between cells, maintain the structural integrity of the stack, and manage heat \cite{Ayers2010,Carmo2013,Bessarabov2016,Feng2017,Irena2020,Parra-Restrepo2020,Doan2021,Teuku2021,Prestat2023,Yasin2024}. These functions must be sustained in the electrolyzer environment, which involves both oxidizing (anode side) and reducing (cathode side) conditions, throughout the electrolyzer’s operational lifespan of 80,000 hours (the U.S. Department of Energy (DOE) ultimate target for 2026). 

\par Titanium is the most commonly used base metal for manufacturing PTLs and BPPs due to its excellent corrosion resistance. However, it is expensive, difficult to machine into complex flow-field designs, and susceptible to hydrogen embrittlement, which can cause mechanical failure. Over time, titanium also develops a thick oxide layer, which increases interfacial contact resistance (ICR) at cell interfaces and thereby reduces electrolyzer efficiency. To address these challenges, precious metal coatings such as Pt or Au are often applied. This significantly increases the cost of already expensive Ti PTLs and BPPs, \cite{Carmo2013,Bessarabov2016,Feng2017,Irena2020,Parra-Restrepo2020,Doan2021,Teuku2021,Prestat2023,Yasin2024,Langemann2015} making up roughly 50--70\% of the total stack costs \cite{Irena2020,Babic2017}.

\par Stainless steel presents a much more affordable and easily machinable alternative to titanium, and it offers substantially higher resistance to hydrogen embrittlement \cite{Carmo2013,Doan2021,Teuku2021,Prestat2023,Yasin2024,Babic2017,Gago2016}. However, in addition to oxide layer formation, dissolved metal cations, primarily $\mathrm{Fe}^{2+}$, can leach into the system and poison the catalyst and electrolyte membrane, leading to increased electrode overpotential and reduced stack lifespan \cite{Babic2017,Sun2014,Papadias2015,Wang2015}.

\par To address oxide formation, metal ion dissolution, and hydrogen embrittlement, a variety of protective coatings have been developed for both Ti and stainless steel components. Thick precious metal coatings are commonly deposited to maintain low ICR, though at high cost; examples include 1~$\upmu\mathrm{m}$ Au by sputtering \cite{Jung2009} and 1.7~$\upmu\mathrm{m}$ Ag-based coatings by pulsed ion plating \cite{Zhang2011} on Ti BPPs; 1~$\upmu\mathrm{m}$ Au by electroplating \cite{Yang2018}, 200~nm Au via PVD \cite{Langemann2015}, or VPS Ti (50--120~$\upmu\mathrm{m}$) plus approximately 1~$\upmu\mathrm{m}$ Pt by sputtering \cite{Gago2016,Lettenmeier2016} on stainless steel BPPs; and 60--180~nm Au \cite{Singh2025,Kang2017a,Kang2017b}, 20--150~nm Ir \cite{Liu2018,Liu2021}, or 200~nm Pt \cite{Rakousky2016,Rakousky2018} by electroplating or sputtering on Ti PTLs.

\par To reduce costs, non-precious metal coatings have been applied, especially for stainless steel components: 50~$\upmu\mathrm{m}$ Ti by VPS plus 1~$\upmu\mathrm{m}$ Nb deposited by sputtering \cite{Lettenmeier2017} on BPPs, 20--50~$\upmu\mathrm{m}$ Nb/Ti by VPS followed by either 1.4~$\upmu\mathrm{m}$ Nb by PVD for BPPs \cite{Stiber2022a} or several~$\upmu\mathrm{m}$ Nb for PTLs \cite{Stiber2022b}; and 30--130~$\upmu\mathrm{m}$ Nb by VPS on Cu BPPs \cite{Kellenberger2022}. As the next step, metal nitride coatings have been tested: TiN/CrN multilayers (9~$\upmu\mathrm{m}$, CPED) \cite{Cheng2023}, TiCr/TiCrN (1.6--3.6~$\upmu\mathrm{m}$, arc ion plating) \cite{Yan2024}, Ta/TaN (0.6--1.2~$\upmu\mathrm{m}$, sputtering) \cite{Ye2024}, and 400~nm NbN by sputtering \cite{Sun2024} on Ti sheets; as well as TiN (0.5~$\upmu\mathrm{m}$, PVD) \cite{Langemann2015} and multilayered Ti/TiN or CrN/TiN ($\sim$2~$\upmu\mathrm{m}$, sputtering) \cite{Rojas2021} on stainless steel sheets. Finally, metal oxide coatings have also been explored, such as ITO/Ta bilayers (1.5~$\upmu\mathrm{m}$ Ta deposited by HiPIMS and 190~nm ITO by PDCMS) on Ti substrates \cite{Laedre2022}, and 50~nm $\mathrm{Ti}{4}\mathrm{O}{7}$ sputtered onto Ti BPPs \cite{Wakayama2021} or the BPP side of Ti PTLs \cite{Wakayama2024}. Another approach enabled ultra-low Pt loadings (<20~nm) by depositing dense, defect-free Nb- and Ti-based coatings via HiPIMS on stainless steel substrates \cite{Varela2024}.

\par While the performance of any coating depends on specific electrolyzer designs and operating conditions, both research and industrial practice indicate that relatively thick Pt overlayers, typically hundreds of nanometers, remain the most reliable approach for metallic BPPs and PTLs to ensure long-term electrolyzer operation. Hence, reducing Pt loading while meeting the DOE target of ICR~<10~$\mathrm{m}\Omega\text{·}\mathrm{cm}^2$ after accelerated corrosion test presents a significant challenge in terms of cost and material efficiency.

\par In this study, we directly address this challenge by developing 200~nm-thick, highly corrosion-resistant, conductive and compact bilayer Ti--Nb--O coatings deposited by HiPIMS on stainless steel sheets, systematically tailoring their composition over a broad range. The optimized coatings enabled DOE-compliant performance while reducing Pt loading by up to two orders of magnitude compared to conventional approaches.

\par Although all coatings were deposited and evaluated on flat stainless steel sheets, the process and properties were developed to enable potential application to both BPPs and PTLs. For Ti PTLs, our coating protocol (including the initial etching step) could be directly implemented to modify contact surfaces on both the BPP and MEA sides. For steel PTLs, however, additional studies are needed to ensure that harmful ions do not leach from uncoated internal surfaces and compromise membrane durability. In such cases, a preliminary treatment to fully cover all internal surfaces, such as Ti electroplating, may be required before applying our coating. By outlining these considerations, we aim to motivate further research to confirm performance on actual PTL architectures and to encourage in situ testing and validation of the presented coating system in operating electrolyzers.

\section{Experimental}\label{sec:level2}

\subsection {Coating preparation}\label{sec:level2a}

\par Coatings were deposited using high-power impulse magnetron sputtering (HiPIMS) technique on substrates consisted of AISI 316L stainless steel sheet (25×50×0.1~$\mathrm{mm}^3$), p-type Si (100) wafer (20×20×0.6~$\mathrm{mm}^3$), and soda-lime glass (75×25×1~$\mathrm{mm}^3$). The substrates were ultrasonically cleaned in isopropyl alcohol for 5 minutes, blow-dried with $\mathrm{N}_{2}$, and placed into a cylindrical vacuum chamber (498~mm in diameter, 403~mm height). A turbomolecular pump (820 l/s for $\mathrm{N}_{2}$, HiPace 800 M, Pfeiffer Vacuum) backed up by a scroll pump ($30 ~ \mathrm{m}^{3}/\mathrm{h}$, ISP-500C, Anest Iwata) ensured a base pressure of $10^{-4} ~ \mathrm{Pa}$. Subsequently, the substrates were heated to 250 °C and held at this temperature during depositions.

\par The depositions were carried out using an unbalanced magnetron with a circular, water-cooled Ti target (100~mm in diameter, 6~mm thick) and a Nb strip (22×15×3~mm) placed in the target racetrack. The Nb content was determined by the sample position relative to the Nb strip fixed on the target; positions 1, 2, 3, and 4 correspond to progressively closer distances to the Nb strip, resulting in increasing Nb content (see Fig. 1). For Ti--O coatings, the Nb strip was omitted. For coatings without Nb, the Nb strip was removed. The substrate-target distance was 100~mm.

\begin{figure}[!h]
\centering
\includegraphics[width=.47\textwidth]{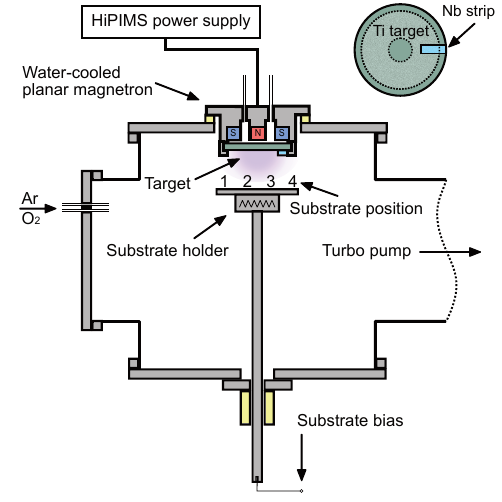}
\caption{\label{fig1} Schematic representation of the experimental setup for coating preparation. Sample positions 1, 2, 3, and 4 correspond to progressively closer distances to the Nb strip. For Ti--O coatings, the Nb strip was omitted.}
\end{figure}

\par Prior to deposition, Ar was introduced at a flow rate of 20 sccm, and the pumping speed of the turbomolecular pump was adjusted using a throttle valve to attain an Ar pressure of 1 Pa. Then, 20 min etching pre-treatment with Ar and metal ions generated by HiPIMS was performed to clean the substrate surface and remove native oxides. This process reduces interface defects, enhances adhesion \cite{Chabanon2022}, and improves both the corrosion resistance and through-plane electrical conductivity of the coated BPP. Both the magnetron and the substrate holder were powered by synchronized pulses from a dual-channel pulse power supply (SIPP2000 USB, MELEC GmbH) equipped with two ADL DC units. The magnetron pulses were 10 µs long and repeated with a frequency of 4000 Hz (corresponding to 4\% duty cycle), delivering an average power of 300 W. The substrate holder received 100 µs pulses at a constant voltage of 700~V, initiated 10 µs prior to each magnetron pulse.

\par For the coating deposition, 50 µs pulses with a repetition frequency of 200 Hz (1\% duty cycle) were applied to the magnetron, delivering an average power of 400 W. The substrates remained at a floating potential. The coatings on the steel and Si substrates comprised two layers, each approximately 100~nm thick. The bottom layer was deposited exclusively in an Ar atmosphere at 1 Pa for each sample to ensure strong adhesion and good electrical contact with the substrate. For the upper layer depositions, oxygen was additionally introduced into the chamber and its flow was precisely regulated using a PID unit (647C, MKS Instruments) to maintain constant oxygen partial pressures, $p_{\mathrm{ox}}$, of 0, 3, 5, or 8~mPa. For X-ray diffraction (XRD), wavelength dispersive spectroscopy (WDS), and electrical resistivity measurements, only the upper layers with a thickness of approximately 500~nm were deposited on the Si and glass substrates for each $p_{\mathrm{ox}}$.

\par Each coated stainless steel sheet was cut in half, with one half subsequently coated with a 5~nm metallic Pt overlayer via RF sputtering. All depositions were performed in a Leybold-Heraeus LH Z400 multitarget system, featuring a stainless steel chamber evacuated by a turbomolecular pump supported by a scroll pump. A Pt target with 99.99\% purity was used, with the target-to-substrate distance maintained at 70~mm and a base pressure below 5~mPa. The target was operated at 15~W, achieving the desired 5~nm thickness after 15~minutes of sputter deposition. To verify the Pt overlayer thickness, a bare Si substrate was included in each run. For this measurement, a small drop of nail polish was applied to the Si substrate prior to deposition and removed afterward, creating a sharp step between coated and uncoated regions. The step height, corresponding to the Pt thickness, was measured using atomic force microscopy (AFM).

\subsection {Elemental composition, structure, and electrical resistivity}\label{sec:level2b}

\par The elemental composition was determined by wavelength dispersive spectroscopy (WDS, MagnaRay, Thermo Fisher Scientific) performed in a scanning electron microscope (SEM, SU-70, Hitachi) using a primary electron energy of 10 keV. For quantitative analysis, a metallic standard of Nb and a rutile standard for Ti and O were used (both Astimex Standards). SEM micrographs were acquired using a secondary electron detector at the same primary energy of electrons. The cross-sectional views were performed on simply broken specimens deposited on pre-scratched Si wafers. 

\par The structure of the as-deposited coatings was characterized by XRD with a PANalytical X’Pert PRO MPD diffractometer operating in the Bragg-Brentano geometry. The measurements were conducted with CuK$\upalpha$ radiation (40 kV, 40 mA), using a 0.25° divergence slit, a 0.5° anti-scatter slit, 0.04 rad Soller slits, a Ni filter to eliminate CuK$\upbeta$ radiation, and an ultrafast X’Celerator semiconductor detector. To minimize strong diffraction signals from the Si(100) substrate, a slightly asymmetrical diffraction geometry with an $\omega$-offset of 1.5° was employed. Scans were performed over a 2$\theta$ range of 8°--108° at a scanning speed of 0.04°/s. The acquired data were processed with the PANalytical HighScore Plus software package.

\par The electrical resistivity of upper layers, approximately 500~nm thick and deposited on glass substrates, was measured using a four-point probe system (Jandel MACOR Probe Head) in a linear configuration.

\subsection {Interfacial contact resistance (ICR) measurement}\label{sec:level2c}

\par
\begin{figure}[h]
\centering
\includegraphics[width=.47\textwidth]{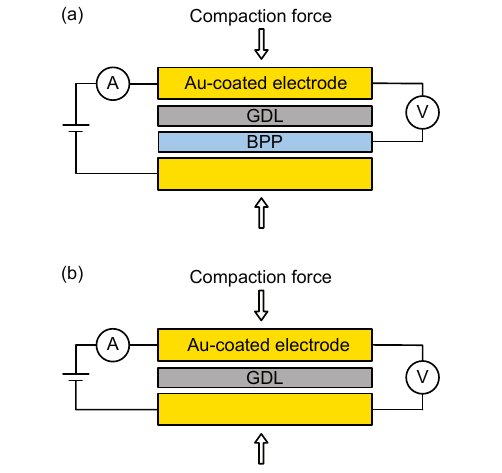}
\caption{\label{fig2} Schematic diagrams of the experimental setups for (a) the ICR measurement and (b) the auxiliary measurement of $R_{\mathrm{Au} / \mathrm{GDL}}$, representing the contact resistance between the gold-coated electrode and the GDL.}
\end{figure}

 ICR measurements were systematically performed both before and after the the electrochemical testing procedure of each sample, adopting an experimental configuration similar to Ref. \cite{Papadias2015}, modified for use with a single side-coated sample. A coated stainless steel sheet sample was stacked with a carbon gas diffusion layer (GDL) and placed between two gold-coated cylindrical copper electrodes with a $1 ~ \mathrm{cm}^{2}$ contact area. Although carbon GDL is unsuitable for the anode side of an operating electrolyzer due to its susceptibility to corrosion, in this study, it was employed for ex-situ testing to ensure comparable ICR values. The GDL used was Toray carbon paper (TGP-H-60) with a thickness of 0.19~mm. A constant direct current of 1 A was applied between the two electrodes using a Keithley 2635B Source Measure Unit, simultaneously measuring the resulting voltage. The described measurement setup, including electrical connections, is schematically represented in Fig.~\ref{fig2} (a). The applied compaction force was progressively increased from 30 to 200 N, corresponding to a compaction pressure range of 0.3 to 2~MPa. 

\par The ICR is defined as follows:
\begin{equation}
\quad \mathrm{ICR}\left[\mathrm{m} \Omega \times \mathrm{cm}^2\right]=R_{\mathrm{GDL} / \mathrm{BPP}}[\mathrm{m} \Omega] \times \mathrm{Area} \left[\mathrm{cm}^2\right]
\end{equation}
where $R_{\mathrm{GDL} / \mathrm{BPP}}$ denotes the contact resistance between the GDL and BPP. This value cannot be measured directly, since the measured resistance contains additional contributions from the measurement setup. All contributions, listed from top to bottom, as depicted in Fig.~\ref{fig2}(a), are given by the following equation:
\begin{equation}
\quad R_{\mathrm{Au} / \mathrm{GDL} / \mathrm{BPP} / \mathrm{Au}}=R_{\mathrm{Au} / \mathrm{GDL}}+R_{\mathrm{GDL}}+R_{\mathrm{GDL} / \mathrm{BPP}}+R_{\mathrm{BPP}}
\end{equation}
where $R_{\mathrm{Au} / \mathrm{GDL}}$ is the contact resistance between the gold-coated electrode and the GDL, which must be determined; $R_{\mathrm{GDL}}$ is the through-plane resistance of the GDL, provided by the manufacturer; $R_{\mathrm{GDL} / \mathrm{BPP}}$ is the aforementioned contact resistance required for the ICR calculation; and $R_{\mathrm{BPP}}$ is the through-plane resistance of the stainless steel sheet substrate (BPP base material), which can be neglected as it is significantly smaller than the other components. To determine $R_{\mathrm{Au} / \mathrm{GDL}}$, an additional measurement was performed with only the GDL placed between the electrodes (see Fig.~\ref{fig2}(b)), comprising the following resistances:
\begin{equation}
\quad R_{\mathrm{Au} / \mathrm{GDL} / \mathrm{Au}}=2 \times R_{\mathrm{Au} / \mathrm{GDL}}+R_{\mathrm{GDL}}
\end{equation}
The ICR was then calculated using the two measured resistances according to the following equation:
\begin{equation}
\quad \mathrm{ICR}=\left(R_{\mathrm{Au} / \mathrm{GDL} / \mathrm{BPP} / \mathrm{Au}}-\frac{R_{\mathrm{Au} / \mathrm{GDL} / \mathrm{Au}}+R_{\mathrm{GDL}}}{2}\right) \times \text { Area }
\end{equation}

\subsection {Electrochemical measurement}\label{sec:level2d}

\par All electrochemical measurements were conducted in a water-jacketed, three-electrode glass corrosion cell filled with 120 mL of $\mathrm{H}_{2}\mathrm{SO}_{4}$ electrolyte solution (pH~5.0) containing 5~ppm NaF, maintained at 60~°C to replicate the operating conditions of a PEM electrolyzer. The samples, serving as the working electrode, were mounted on the PTFE cell wall and sealed with a Viton gasket, exposing a circular area of $1 ~ \mathrm{cm}^{2}$ to the electrolyte. An Ag/AgCl (3M KCl) reference electrode was positioned near the working electrode via a Luggin capillary to minimize potential drop, while a Pt wire functioned as the counter electrode. All electrodes were connected to a potentiostat (Squidstat Plus, Admiral Instruments). Accelerated corrosion test consisted of potentiostatic polarization at 2~V vs.~SHE (equivalent to 2.3~V vs.~RHE) for 1 hour. Potentiodynamic polarization measurements (0--2~V vs.~SHE) were performed at a scan rate of 1~mV/s, both before and after the potentiostatic polarization.

\section{Results}\label{sec:level3}

\subsection {Elemental composition and structure}\label{sec:level3a}

\par Figure 3 presents a ternary plot illustrating the elemental compositions of the upper layers of all prepared coatings. The upper layers of the coatings were synthesized under $p_{\mathrm{ox}}$ = 0, 3, 5, and 8~mPa, denoted by black triangles, red diamonds, green circles, and purple squares, respectively. The Nb content in the coatings ranged from 0 to 10.5~at.\%, corresponding to $\mathrm{Nb}/(\mathrm{Ti} + \mathrm{Nb})$ ratios of 0 to approximately 0.14. This variation indicates a controlled incorporation of Nb into the Ti or Ti–O matrix, enabling precise tuning of the coating composition to achieve the desired properties.

\begin{figure}[h]
\centering
\includegraphics[width=.47\textwidth]{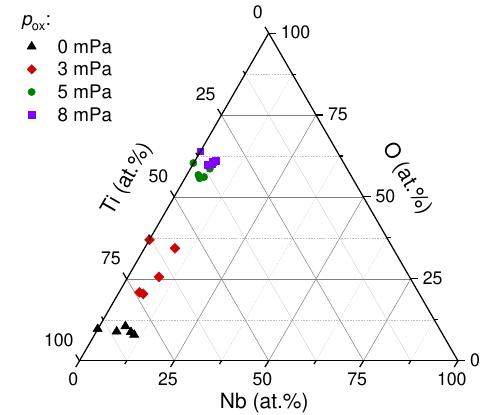}
\caption{\label{fig3} The ternary plot depicts the elemental compositions of the upper layers for all prepared coatings. Black triangles, red diamonds, green circles, and purple squares represent layers prepared at oxygen partial pressures, $p_{\mathrm{ox}}$, of 0, 3, 5, and 8~mPa, respectively.}
\end{figure}

\par The O content in the coatings exhibited a pronounced dependence on $p_{\mathrm{ox}}$ during the deposition. At $p_{\mathrm{ox}} = 0~\mathrm{mPa}$, the coatings contained approximately 9~at.\% O, resulting in an $\mathrm{O}/(\mathrm{Ti} + \mathrm{Nb})$ ratio of about 0.1. This minimal O incorporation was primarily attributed to the desorption of residual O from the chamber’s inner surfaces. Upon introducing oxygen at $p_{\mathrm{ox}} = 3~\mathrm{mPa}$, the O content increased to 20--37~at.\%, with an $\mathrm{O}/(\mathrm{Ti} + \mathrm{Nb})$ ratio ranging from 0.26 to 0.6. At $p_{\mathrm{ox}} = 5~\mathrm{mPa}$, the O content further rose to 56--60~at.\%, corresponding to $\mathrm{O}/(\mathrm{Ti} + \mathrm{Nb})$ ratios of 1.3--1.5. At the maximum $p_{\mathrm{ox}}$ of 8~mPa, the coatings exhibited an O content of 60--64~at.\%, with $\mathrm{O}/(\mathrm{Ti} + \mathrm{Nb})$ ratios between 1.5 and 1.8. For $p_{\mathrm{ox}}$ = 3, 5, and 8~mPa, the highest O contents were observed for coatings without Nb, while lower values were found at low or intermediate Nb contents. Overall, these results demonstrate that the O incorporation increases markedly with $p_{\mathrm{ox}}$, indicating the formation of substantial Ti--O and Nb--O bonds in the coatings. Furthermore, the Nb content notably influences the amount of O incorporation, particularly at $p_{\mathrm{ox}} = 3~\mathrm{mPa}$.

\begin{figure}[h]
\centering
\includegraphics[width=.47\textwidth]{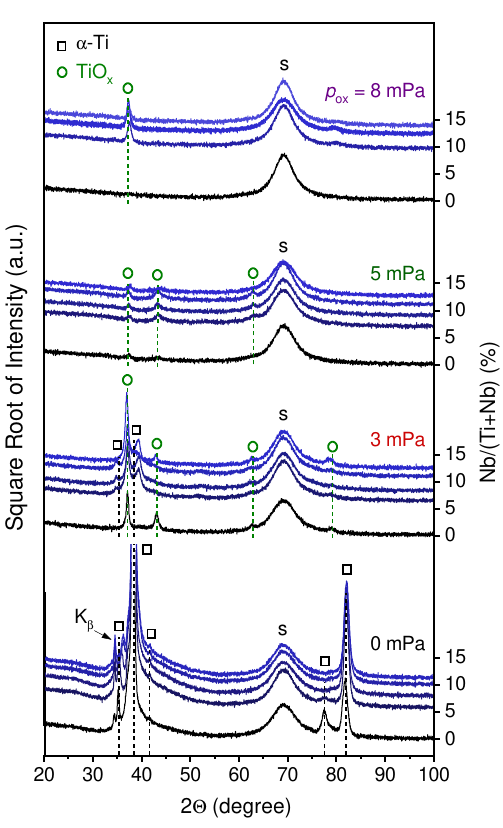}
\caption{\label{fig4} X-ray diffraction (XRD) patterns of the upper layers within the examined compositional space. The patterns are grouped according to the oxygen partial pressure, $p_{\mathrm{ox}}$, applied during deposition (0, 3, 5, and 8~mPa) and are arranged vertically based on the Nb content in the metal fraction, expressed as $\mathrm{Nb}/(\mathrm{Ti} + \mathrm{Nb})$. Reflections of the hexagonal $\alpha$-Ti phase are denoted by black squares, while those of the substoichiometric cubic TiO$_x$ phase are marked by green circles. The Si substrate peak is denoted by ``s''.}
\end{figure}

\par XRD analysis of the upper layers of the coatings deposited at $p_{\mathrm{ox}} = 0~\mathrm{mPa}$ revealed that the hexagonal $\alpha$-Ti phase (PDF card no. 00-044-1294) was dominant across all investigated Nb contents, exhibiting a pronounced (0002) preferential orientation. No distinct reflections attributable to the cubic $\beta$-Ti phase (PDF no. card 00-044-1288) were observed in the diffraction patterns. However, the main $\beta$-Ti peak overlapped with the most intense $\alpha$-Ti peak, which could potentially mask its presence. As the Nb content increases, the likelihood of the $\beta$-Ti phase formation rises, since Nb forms a solid solution with $\beta$-Ti due to their common BCC lattice structure. On the other hand, the solubility limit of Nb in $\alpha$-Ti, which is 2~at.\% according to the equilibrium phase diagram, could be increased in our case due to the non-equilibrium conditions of our deposition process. This fact is supported by a gradual shift of the $\alpha$-Ti $(1~0~\overline{1}~0)$ peak to higher diffraction angles (from 35.2$^\circ$ to 36.2$^\circ$) in Fig. 4 for $p_{\mathrm{ox}} = 0~\mathrm{mPa}$. Since other different peaks remained at the same positions with Nb incorporation, the shift of the $(10\overline{1}~0)$ peak indicates that Nb incorporation induces anisotropic lattice distortion, resulting in a contraction of the $a$-axis in the hexagonal $\alpha$-Ti lattice.

\par The O introduction at $p_{\mathrm{ox}}$ values between 3 and 8~mPa significantly influenced the phase composition of the upper layers of the coatings. While the $\alpha$-Ti phase can dissolve up to approx. 33~at.\% of O~\cite{Okamoto2011}, the addition of $p_{\mathrm{ox}} = 3~\mathrm{mPa}$ led to the formation of a substoichiometric cubic TiO$_x$ phase, although peaks corresponding to the $\alpha$-Ti phase remained still detectable, especially for the Ti--Nb--O coatings. As $p_{\mathrm{ox}}$ increased to 5 and 8~mPa, the phase composition changed further, and only the cubic TiO$_x$ phase was observed. The $\mathrm{O}/(\mathrm{Ti} + \mathrm{Nb})$ ratio in this phase varied from 0.6 to 1.3 (e.g., PDF cards no. 04-005-4341 and 04-005-4348), reflecting its substoichiometric existence over a broad range. Additionally, the presence of an amorphous background around 30$^\circ$ became more pronounced with increasing O incorporation. The results further indicate that Nb was most likely substitutionally dissolved within the crystal lattices of both the $\alpha$-Ti and TiO$_x$ phases, rather than forming separate Nb-rich phases. Collectively, these findings underscore the crucial role of $p_{\mathrm{ox}}$ in determining both the phase composition and structural properties of the prepared coatings.

\begin{figure}[h]
\centering
\includegraphics[width=.47\textwidth]{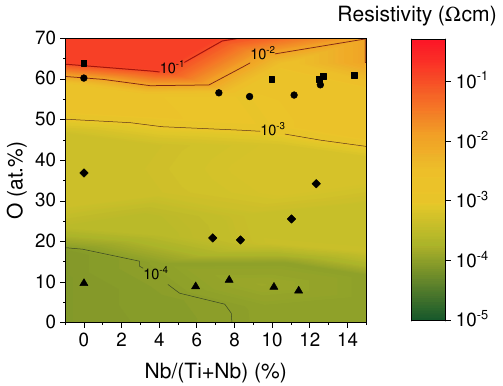}
\caption{\label{fig5} Electrical resistivity of coating upper layers visualized through a continuous color mapping within the examined compositional space. Data smoothing is applied to highlight overall trends within each panel. Triangle, diamond, circle, and square symbols indicate the elemental compositions of coatings with upper layers deposited at $p_{\mathrm{ox}}$ values of 0, 3, 5, and 8~mPa, respectively.}
\end{figure}

\par Figure~5 shows that the electrical resistivity of the upper layers of the coatings is strongly influenced by the O content. This dependence is much more pronounced than that on the Nb content in the metallic fraction, expressed as $\mathrm{Nb}/(\mathrm{Ti} + \mathrm{Nb})$. The observed trend closely parallels the changes in the crystalline structure revealed by XRD, where phase formation was primarily governed by $p_{\mathrm{ox}}$ during the deposition. These findings indicate that O availability played a dominant role in determining both the electrical and structural properties of the coatings.

\par The lowest electrical resistivity was measured for the Ti--O and Ti--Nb--O upper layers deposited at $p_{\mathrm{ox}} = 0~\mathrm{mPa}$, ranging from $0.9\times10^{-4}$ to $1.3\times10^{-4}$~$\Omega\text{·}\mathrm{cm}$, with the lowest value observed for the Ti--O layer. For the upper layers prepared at $p_{\mathrm{ox}} = 3~\mathrm{mPa}$, the resistivity remained relatively low, ranging from $2.7\times10^{-4}$ to $6.1\times10^{-4}$~$\Omega\text{·}\mathrm{cm}$; the lowest values corresponded to layers with lower Nb doping ($\mathrm{Nb}/(\mathrm{Ti} + \mathrm{Nb})$ of 6.8 and 8.3~at.\%). Further increases in $p_{\mathrm{ox}}$ to 5 and 8~mPa caused the resistivity to rise by approximately one order of magnitude, reaching $1.3 \times 10^{-3}$ to $1.0\times 10^{-2}~\Omega\text{·}\mathrm{cm}$. An exception was observed for the layer prepared at 8~mPa without Nb, where the resistivity was substantially higher ($1.4 \times 10^{-1}$~$\Omega\text{·}\mathrm{cm}$).

\par The bilayer structure of the coating is clearly visible in the cross-sectional SEM image (Fig. 6). The bottom layer deposited at $p_{\mathrm{ox}} = 0~\mathrm{mPa}$ appears to be more compact, while the upper layer deposited at $p_{\mathrm{ox}} = 5~\mathrm{mPa}$ most likely exhibits larger columnar grains due to the O addition during the deposition. Analysis of the top-view SEM image (inset of Fig. 6) indicates that the grains in the upper layer have an average lateral diameter of approximately 20~nm. Overall, the coating structure observed in the cross section appears relatively dense, which is advantageous for enhancing corrosion resistance.

\begin{figure}[h]
\centering
\includegraphics[width=.47\textwidth]{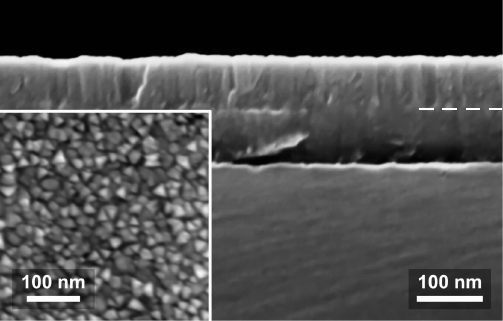}
\caption{\label{fig6} Cross-sectional SEM image of Ti--O coating deposited on a silicon substrate. The bottom layer was deposited at $p_{\mathrm{ox}} = 0~\mathrm{mPa}$, while the upper layer was deposited at $p_{\mathrm{ox}} = 5~\mathrm{mPa}$. The inset shows a top-view SEM image of the coating.}
\end{figure}

\subsection {Electrochemical and ICR measurements}\label{sec:level3b}

\par To assess the intrinsic properties of the coatings, potentiodynamic polarization and ICR measurements were performed for all studied compositions prior to Pt overlayer deposition, both before and after accelerated corrosion test, which consisted of potentiostatic polarization at 2~V vs. SHE (equivalent to 2.3~V vs. RHE) for 1~hour. Key properties such as corrosion current densities, $J_\mathrm{corr}$, and corrosion potentials, $E_\mathrm{corr}$, were determined from the Tafel plots constructed using the potentiodynamic polarization data (representative examples are shown in Fig.~8b). The polarization resistance, $R_{\mathrm{p}}$, for each sample was also calculated from the anodic and cathodic slopes according to the Stern--Geary equation. The complete set of $J_\mathrm{corr}$, $E_\mathrm{corr}$, $R_{\mathrm{p}}$, and ICR data for all coatings, both before and after the potentiostatic polarization, is provided in Supplementary Table~1a. It should be noted that, although the US Department of Energy (DOE) specifies target values for $J_\mathrm{corr}$ below 1~$\upmu\mathrm{A}/\mathrm{cm}^2$ and ICR below 10~$\mathrm{m}\Omega\text{·}\mathrm{cm}^2$ for bipolar plates in PEM fuel cells, these benchmarks are widely used as a reference to evaluate the performance of coatings for PEM electrolyzers as well.

\par All coatings demonstrated exceptional corrosion resistance, with $J_\mathrm{corr}$ values in the range of $0.01$--$0.06$~$\upmu\mathrm{A}/\mathrm{cm}^2$ before and $0.01$--$0.08$~$\upmu\mathrm{A}/\mathrm{cm}^2$ after the potentiostatic polarization (Fig.~7a and~b). Ti--Nb--O coatings with the upper layer prepared at $p_{\mathrm{ox}} = 3~\mathrm{mPa}$ exhibited moderately higher $J_\mathrm{corr}$ compared to the rest. Interestingly, after the potentiostatic polarization, the $J_\mathrm{corr}$ of these coatings decreased by 50--80\%. This decrease can be attributed to surface passivation during the potentiostatic polarization, which enhances the protective properties of the coatings. In contrast, coatings with upper layers deposited at $p_\mathrm{ox}$ of 5 or 8~mPa showed an increase in $J_\mathrm{corr}$ (up to 350\%) most likely caused by the development of voids or other defects during the potentiostatic polarization. This may be attributed to a higher amorphous fraction and disorder in the structure of the oxidized upper layer, as suggested by the XRD results shown in Fig.~4, which can result in less effective barrier formation. Nonetheless, even the highest $J_\mathrm{corr}$ values measured are more than one order of magnitude lower than the U.S.\ DOE criterion.

\par While $J_{\mathrm{corr}}$ remains the key metric for evaluating corrosion performance, the corrosion potential, $E_\mathrm{corr}$, and polarization resistance, $R_{\mathrm{p}}$, also provide valuable insights. High $E_\mathrm{corr}$ values indicate effective passivation with increased resistance to oxidative reactions, while high $R_{\mathrm{p}}$ values reflect not only effective passivation but also a compact, low-porosity, defect-free surface—both crucial for minimizing metal ion dissolution and ensuring long-term stability in PEM electrolyzer stacks.

\par Before the potentiostatic polarization, $E_\mathrm{corr}$ ranged from $-0.2$ to $-0.03$~V vs.~SHE and tended to increase for higher $p_{\mathrm{ox}}$, suggesting that higher O content leads to a more passivated, albeit less conductive surface. This trend is consistent with the $J_\mathrm{corr}$ results (see Fig.~7a), where higher $p_{\mathrm{ox}}$ corresponds to lower $J_\mathrm{corr}$. Regarding $R_{\mathrm{p}}$, the values ranged from $0.5$ to $2.6~\times 10^6$~$\Omega\text{·}\mathrm{cm}^2$, closely mirroring the trend observed for $J_\mathrm{corr}$. Notably, these $R_{\mathrm{p}}$ values are substantially higher than those typically reported in the literature, such as $4.4 \times 10^5$~$\Omega\text{·}\mathrm{cm}^2$ for Nb~\cite{Kellenberger2020}, $8 \times 10^4$~$\Omega\text{·}\mathrm{cm}^2$ for TiN~\cite{Li2021}, and $5 \times 10^4$~$\Omega\text{·}\mathrm{cm}^2$ for Ta/TaN~\cite{Ye2024}. This improvement may be attributed to the smooth and compact morphology of the coatings deposited by HiPIMS, as well as differences in the parameters of the accelerated corrosion test, such as electrolyte composition, temperature, and other experimental conditions.

\begin{figure*}[t]
\centering
\includegraphics[width=0.94\textwidth]{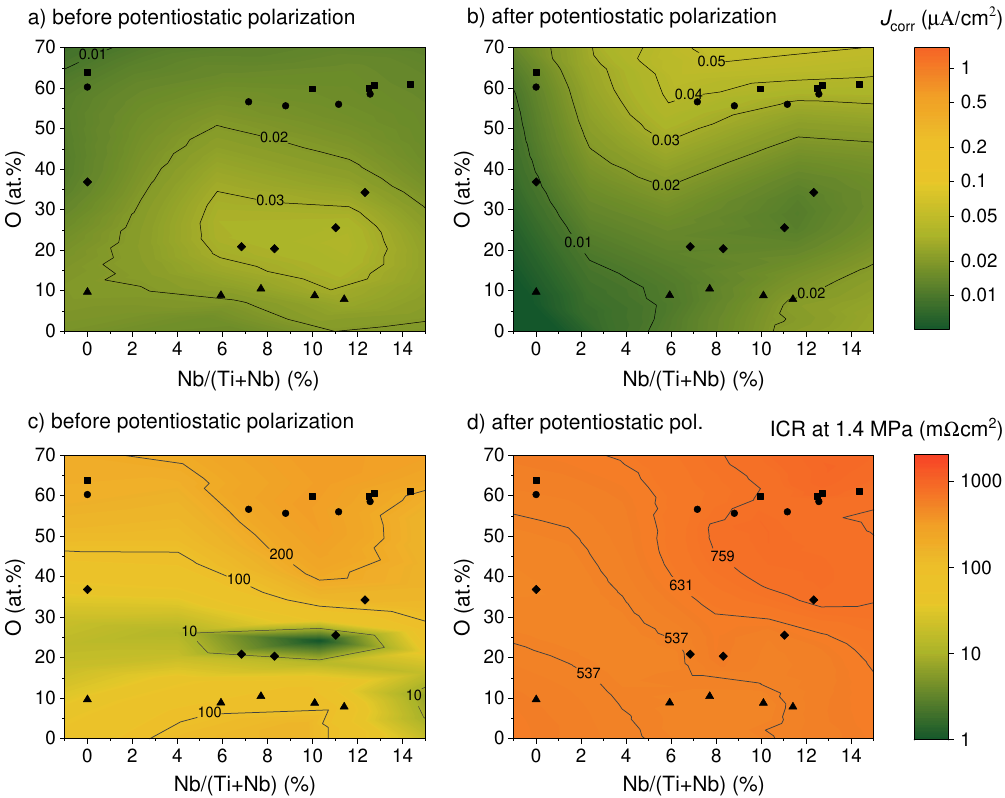}
\caption{\label{fig7} Potentiodynamic polarization corrosion current density ($J_\mathrm{corr}$) and interfacial contact resistance (ICR) of coatings deposited on stainless steel substrates under various conditions. The indicated Nb content in the metal fraction, $\mathrm{Nb}/(\mathrm{Ti} + \mathrm{Nb})$, and O content apply only to the upper layer, which is in direct contact with the electrolyte. Panels (a) and (c) show $J_\mathrm{corr}$ and ICR measured before the potentiostatic polarization, while panels (b) and (d) present the values after the potentiostatic polarization. The values are visualized using continuous color mapping across the compositional space of the upper layer. Data smoothing is applied to highlight overall trends within each panel. Triangle, diamond, circle, and square symbols indicate the elemental compositions of coatings with upper layers deposited at $p_{\mathrm{ox}}$ values of 0, 3, 5, and 8~mPa, respectively. The complete set of data is provided in Supplementary Table~1a.}The complete set of data is provided in Supplementary Table~1a.
\end{figure*}

\par Following the potentiostatic polarization, $E_\mathrm{corr}$ for all coatings converged to a narrow range between $-0.06$ and $0$~V vs.~SHE, with the highest values observed for Ti--Nb--O coatings with a high Nb content and upper layers deposited at $p_{\mathrm{ox}} = 3~\mathrm{mPa}$. This convergence indicates a comparable degree of surface passivation across the compositional range. In contrast, $R_{\mathrm{p}}$ continued to correlate with $J_{\mathrm{corr}}$ after the potentiostatic polarization (see Fig.~7b). This is likely because both parameters are sensitive to variations in surface porosity and defect formation, whereas $E_\mathrm{corr}$ does not directly reflect changes in surface area. All Ti--O coatings exhibited only a slight increase in $R_{\mathrm{p}}$, while Ti--Nb--O coatings with upper layers deposited at $p_{\mathrm{ox}} = 3~\mathrm{mPa}$ showed a substantially greater increase following the potentiostatic polarization, reaching up to $4.3 \times 10^6$~$\Omega\text{·}\mathrm{cm}^2$ for $p_{\mathrm{ox}} = 3~\mathrm{mPa}$ and the second-highest Nb content (where the Nb content and the $\mathrm{Nb}/(\mathrm{Ti} + \mathrm{Nb})$ ratio in the Ti--Nb--O upper layer is 8.2~at.\% and 0.11, respectively). Notably, this coating also exhibited the lowest ICR prior to the potentiostatic polarization. This improvement in $R_{\mathrm{p}}$ suggests enhanced passivation due to the formation of a compact, defect-free passive oxide layer. Conversely, Ti--Nb--O coatings with upper layers deposited at higher $p_{\mathrm{ox}}$ values (5 and 8~mPa) exhibited a slight decrease in $R_{\mathrm{p}}$ after polarization, with values ranging from $5 \times 10^5$ to $1 \times 10^6$~$\Omega\text{·}\mathrm{cm}^2$. This reduction was likely caused by an increase in surface defects and the development of voids, consistent with the observed increase in $J_\mathrm{corr}$ for these coatings.

\par For the ICR before the potentiostatic polarization, most coatings with upper layers deposited at $p_{\mathrm{ox}}$ of 0 and 3~mPa exhibited ICR values below 100~$\mathrm{m}\Omega\text{·}\mathrm{cm}^2$. Notably, coatings with intermediate Nb content and an upper layer deposited at $p_{\mathrm{ox}} = 3~\mathrm{mPa}$ demonstrated the lowest initial ICR, falling below 7~$\mathrm{m}\Omega\text{·}\mathrm{cm}^2$. These results correlate with the higher $J_\mathrm{corr}$ values before the potentiostatic polarization for these coatings, indicating higher surface conductivity. This can be attributed to a substoichiometric cubic TiO$_x$ phase with high conductivity in the upper layer deposited at $p_{\mathrm{ox}} = 3~\mathrm{mPa}$ (see Fig.~4). Unlike the metallic $\alpha$-Ti phase, which readily forms a surface-passivating non-conductive TiO$_2$ oxide layer upon exposure to air, the substoichiometric TiO$_x$ phase tends to incorporate oxygen into its bulk to increase its stoichiometry, suppressing resistive surface TiO$_2$ layer formation and maintaining lower ICR. In contrast, coatings with upper layers deposited at $p_{\mathrm{ox}}$ of 5 and 8~mPa generally showed ICR values above 200~$\mathrm{m}\Omega\text{·}\mathrm{cm}^2$ (Fig.~7c), as the increased oxygen content in the substoichiometric TiO$_x$ phase in the upper layer leads to its higher resistivity (see Fig.~5), likely impeding efficient electron transfer at the contact with the GDL wires.

\par After the potentiostatic polarization, ICR values for all coatings increased substantially, exceeding 500~$\mathrm{m}\Omega\text{·}\mathrm{cm}^2$ (Fig.~7d), due to increased surface oxidation that impeded effective contact with the GDL. This is consistent with the observed increase in $E_\mathrm{corr}$ for all coatings, which converged to a narrow range. Notably, ICR values tended to increase with both higher O and Nb content in the upper layers, reaching a maximum of 900~$\mathrm{m}\Omega\text{·}\mathrm{cm}^2$, which again correlates with the $J_\mathrm{corr}$ trend observed after the potentiostatic polarization.

\par \par In summary, fine-tuning the composition revealed the optimal parameters for coating performance: a lower oxygen partial pressure, $p_\mathrm{ox} = 3$~mPa, where the O content and the $\mathrm{O}/(\mathrm{Ti} + \mathrm{Nb})$ ratio in the Ti--Nb--O upper layer range from 21 to 26~at.\% and 0.26 to 0.34, respectively, and an intermediate Nb content, where the Nb content and the $\mathrm{Nb}/(\mathrm{Ti} + \mathrm{Nb})$ ratio in the upper layer range from 5.4 to 8.2~at.\% and 0.07 to 0.11, respectively. These coatings exhibited the lowest initial ICR and, after the potentiostatic polarization, showed reduced $J_\mathrm{corr}$ as well as elevated $E_\mathrm{corr}$ and $R_\mathrm{p}$ values, indicating effective surface passivation while maintaining a compact and defect-free surface.

\subsection {Electrochemical and ICR measurements after Pt overlayer application}\label{sec:level3c}

\par Since none of the coatings met the DOE requirement of maintaining ICR after the potentiostatic polarization below $10~\mathrm{m}\Omega\text{·}\mathrm{cm}^2$, we first investigated the dependence of ICR on Pt layer thickness before and after the potentiostatic polarization (Fig.~8a). We selected a Ti--O coating with the upper layer deposited at $p_{\mathrm{ox}} = 3~\mathrm{mPa}$. This coating exhibited low and stable $J_\mathrm{corr}$ values both prior to and after the potentiostatic polarization, with a relatively low initial ICR of $37.5~\mathrm{m}\Omega\text{·}\mathrm{cm}^2$. The addition of only $2~\mathrm{nm}$ of Pt reduced the initial ICR to $25.2~\mathrm{m}\Omega\text{·}\mathrm{cm}^2$ and led to a substantial decrease in post-corrosion ICR, from $496$ to $86.5~\mathrm{m}\Omega\text{·}\mathrm{cm}^2$. Moreover, a $5~\mathrm{nm}$ Pt overlayer was already sufficient to meet the DOE target, achieving ICR values of $7.6~\mathrm{m}\Omega\text{·}\mathrm{cm}^2$ before and $5.6~\mathrm{m}\Omega\text{·}\mathrm{cm}^2$ after the potentiostatic polarization. Increasing the Pt thickness to $20~\mathrm{nm}$ did not yield any further improvement in ICR. 

\par Based on these findings, all coatings were subsequently coated with a $5~\mathrm{nm}$ Pt overlayer. Both ICR and potentiodynamic measurements were performed before and after the potentiostatic polarization, and the resulting $J_\mathrm{corr}$ and ICR values are presented in Fig.~9. The complete set of $J_\mathrm{corr}$, $E_\mathrm{corr}$, $R_{\mathrm{p}}$, and ICR data for all coatings with a 5~nm Pt overlayer before and after the potentiostatic polarization is provided in Supplementary Table~1b.

\begin{figure}[t]
\centering
\includegraphics[width=.47\textwidth]{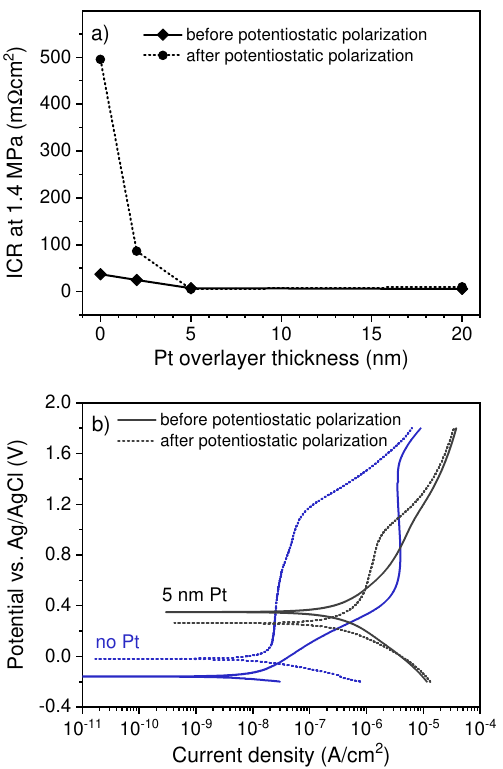}
\caption{\label{fig8} (a) Interfacial contact resistance (ICR) of Ti--O coating deposited on stainless steel substrate, with the upper layer deposited at $p_{\mathrm{ox}} = 3~\mathrm{mPa}$, as a function of Pt overlayer thickness, measured before and after the potentiostatic polarization. (b) Tafel plot from potentiodynamic measurements before and after the potentiostatic polarization for the coating shown in panel (a), with and without a 5~nm Pt overlayer.}
\end{figure}

\begin{figure*}[t]
\centering
\includegraphics[width=0.94\textwidth]{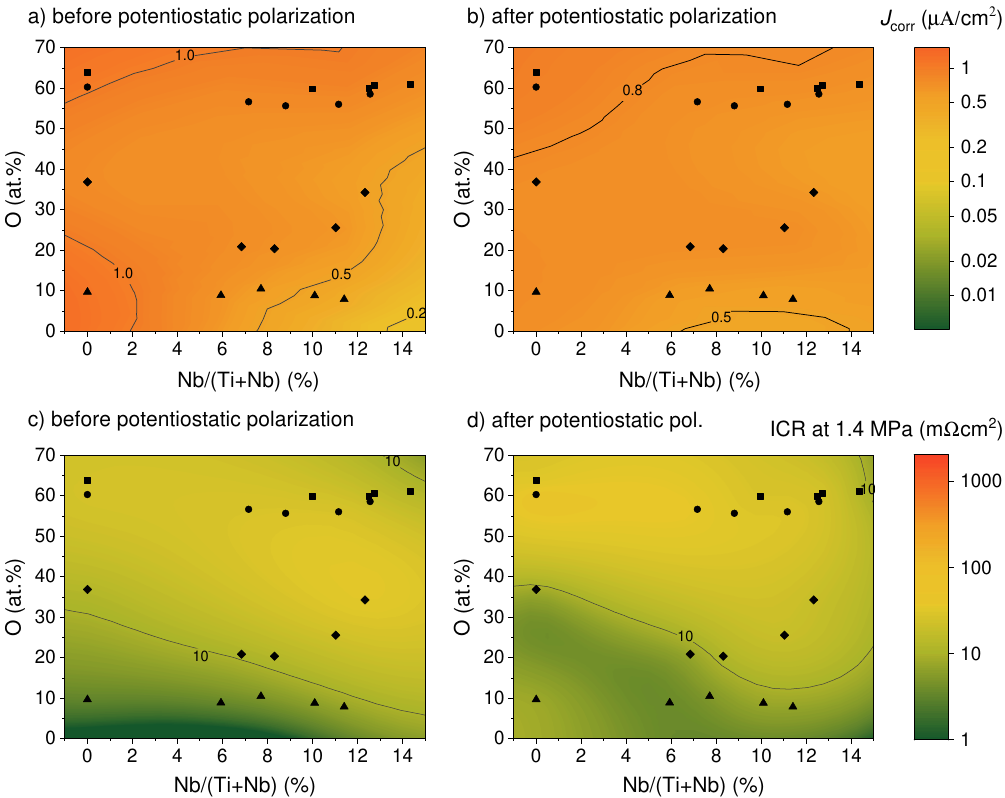}
\caption{\label{fig9} Potentiodynamic polarization corrosion current density ($J_\mathrm{corr}$) and interfacial contact resistance (ICR) of coatings deposited on stainless steel substrates under various conditions with a 5~nm Pt overlayer. The indicated Nb content in the metal fraction, $\mathrm{Nb}/(\mathrm{Ti} + \mathrm{Nb})$, and O content apply only to the upper layer, which is in direct contact with the electrolyte. Panels (a) and (c) show $J_\mathrm{corr}$ and ICR measured before the potentiostatic polarization, while panels (b) and (d) present the values after the potentiostatic polarization. The values are visualized using continuous color mapping across the compositional space of the upper layer. Data smoothing is applied to highlight overall trends within each panel. Triangle, diamond, circle, and square symbols indicate the elemental compositions of coatings with upper layers deposited at $p_{\mathrm{ox}}$ values of 0, 3, 5, and 8~mPa, respectively. The complete set of data is provided in Supplementary Table~1b.}
\end{figure*}

\par Tafel plot in Fig.~8b shows that both Ti--O coatings, with and without a Pt overlayer, exhibited enhanced passivation behavior after the potentiostatic polarization. For the Ti--O coating without Pt, the onset of passivation was already evident before polarization; however, after polarization, the passivation region began at a lower potential and the current density was significantly reduced across the entire anodic potential range. The Pt-coated sample showed no clear passivation before polarization, but developed a passive region afterward, likely due to discontinuities in the thin (5~nm) Pt layer that exposed the underlying Ti--O to passivation.

\par The $J_\mathrm{corr}$ values for all coatings with a 5~nm Pt overlayer were in the range of $0.3$--$1.3$~$\upmu\mathrm{A}/\mathrm{cm}^2$ before the potentiostatic polarization (Fig.~9a), with the highest value observed for the Ti--O coating deposited at $p_{\mathrm{ox}} = 0~\mathrm{mPa}$. After the potentiostatic polarization, $J_\mathrm{corr}$ values decreased slightly to a range of $0.2$--$1.0$~$\upmu\mathrm{A}/\mathrm{cm}^2$ (Fig.~9b), satisfying the DOE requirement. A slight decreasing trend in $J_\mathrm{corr}$ with increasing Nb content was observed. It should be noted that these $J_\mathrm{corr}$ values are approximately one order of magnitude higher than those measured for coatings without a Pt overlayer (see, e.g., Fig.~8b), which can be attributed to the high surface conductivity imparted by the Pt, maintained even after the potentiostatic polarization.

\par The $E_\mathrm{corr}$ values ranged from $0.29$ to $0.38$~V vs.~SHE before the potentiostatic polarization, and decreased to $0.20$ to $0.31$~V vs.~SHE after the potentiostatic polarization (see, e.g., Fig.~8b). These values are significantly higher than those of coatings without a Pt overlayer, indicating enhanced resistance to oxidative reactions, including metal dissolution.

\par The $R_{\mathrm{p}}$ values prior to the potentiostatic polarization were fairly consistent across all studied compositions in the range from $9 \times 10^4$ to $2.8 \times 10^5$~$\Omega\text{·}\mathrm{cm}^2$, with the lowest value observed for the Ti--O coating deposited at $p_{\mathrm{ox}} = 0~\mathrm{mPa}$. After the potentiostatic polarization, $R_{\mathrm{p}}$ values for all coatings slightly increased to a range of $1.2$--$4.0 \times 10^5$~$\Omega\text{·}\mathrm{cm}^2$. These $R_{\mathrm{p}}$ values measured for the coatings with Pt overlayer are approximately an order of magnitude lower than the values observed for coatings without Pt overlayer both before and after the potentiostatic polarization, reflecting the ability of the Pt overlayer to effectively preserve its high conductivity throughout the potentiostatic polarization.

\par The ICR values of all coatings measured before the potentiostatic polarization dropped below $50~\mathrm{m}\Omega\text{·}\mathrm{cm}^2$ after being coated with a 5~nm Pt overlayer (Fig.~9c). This substantial decrease is not due to a reduction in the bulk resistivity of the coatings, but is primarily attributed to enhanced electron transfer between the GDL wires and the upper layer. The Pt overlayer provides a metallic interface with the GDL wires and increases the effective contact area with the upper layer. This effect is particularly significant for upper layers with higher initial electrical resistivity, where electron transfer at the contact is otherwise limited.

\par The lowest pre-corrosion ICR values were observed for Ti--O and Ti--Nb--O coatings deposited at $p_{\mathrm{ox}} = 0~\mathrm{mPa}$, with values below $5~\mathrm{m}\Omega\text{·}\mathrm{cm}^2$, except for the coating with the highest Nb content. Unexpectedly, Ti--Nb--O coatings with the highest and second-highest Nb contents deposited at $p_{\mathrm{ox}} = 8~\mathrm{mPa}$ also exhibited ICR values below $5~\mathrm{m}\Omega\text{·}\mathrm{cm}^2$, deviating from the general trend. Some deviations in ICR values after Pt deposition, measured prior to potentiostatic polarization, may arise from incomplete or non-uniform Pt coverage, especially on rough or porous surfaces. Notably, industry-relevant deposition of the Pt overlayer in a single uninterrupted process (without exposing the upper layer to air) would likely yield even lower and more consistent ICR values prior to polarization, as this approach minimizes interfacial oxidation and contamination.

\par The most relevant results for the long-term stability and durability of the coatings under operational conditions are those obtained after the potentiostatic polarization test. As shown in Fig.~9d, ICR values remained largely stable, with only minor variations. For several compositions, a slight decrease in ICR was observed after polarization, particularly for Ti--Nb--O coating deposited at $p_{\mathrm{ox}} = 0~\mathrm{mPa}$ with the lowest Nb content (from 4.9 to $4~\mathrm{m}\Omega\text{·}\mathrm{cm}^2$), for the Ti--O coating with upper layer deposited at $p_{\mathrm{ox}} = 3~\mathrm{mPa}$ (from 7.6 to $5.6~\mathrm{m}\Omega\text{·}\mathrm{cm}^2$), and for Ti--Nb--O coating deposited at $p_{\mathrm{ox}} = 0~\mathrm{mPa}$ with the lowest Nb content and upper layer deposited at $p_{\mathrm{ox}} = 3~\mathrm{mPa}$ (from 7.3 to $5.3~\mathrm{m}\Omega\text{·}\mathrm{cm}^2$). Notably, these coatings also exhibited some of the lowest ICR values after the potentiostatic polarization without the Pt overlayer (see Fig.~7d). For all remaining coatings, the ICR either remained similar or increased after the potentiostatic polarization.

\par Although the 5~nm Pt overlayer provides substantial initial ICR improvement, it only partially limits oxygen diffusion and does not fully prevent oxygen penetration into the underlying coating. Therefore, the intrinsic properties of the bare coating without the Pt overlayer---determined by its composition and structure, as discussed in Sec.~3.2---remain critical for long-term resistance to passivation and degradation, ensuring durable low ICR and sustained performance during operation.

\par
Overall, these findings demonstrate that the developed coatings enable the DOE ICR target to be met after accelerated corrosion testing with a Pt overlayer as thin as 5~nm, resulting in a significant reduction in Pt loading.

\section{\label{sec:level4}Conclusions}

\par In this study, we demonstrate that precise tailoring of both composition and structure in 200~nm-thick bilayer Ti--Nb--O coatings deposited by HiPIMS enables the simultaneous achievement of outstanding corrosion resistance and high electrical conductivity. By optimizing the O and Nb content, the coatings maintain the compact structure after electrochemical tests and allow a significant reduction in Pt loading, which is, however, still needed to meet U.S.~DOE requirements.

\par The incorporation of Nb into the coatings contributed to reducing resistivity, with optimized compositions reaching values on the order of $10^{-4}~\Omega\text{·}\mathrm{cm}$. The Ti--Nb--O coatings exhibited excellent corrosion resistance, with extremely low corrosion current densities $J_\mathrm{corr} = 0.01$--$0.08$~$\upmu\mathrm{A}/\mathrm{cm}^2$ measured after accelerated corrosion test, consisting of potentiostatic polarization at 2~V vs. SHE (equivalent to 2.3~V vs. RHE) for 1~hour, yielding values more than an order of magnitude below the U.S. DOE criterion. Polarization resistance, $R_\mathrm{p}$, reached up to $4.3 \times 10^6~\Omega\text{·}\mathrm{cm}^2$, confirming the preservation of compact and smooth surfaces even after the potentiostatic polarization for certain compositions.

\par Fine-tuning the Ti--Nb--O upper layer composition revealed that a lower oxygen partial pressure, $p_\mathrm{ox} = 3$~mPa, resulting in an O content of 21 to 26~at.\%, combined with intermediate Nb doping, with an Nb content of 5.4 to 8.2~at.\%, led to optimal coating performance. These coatings exhibited the lowest initial ICR, and after the potentiostatic polarization, showed reduced $J_\mathrm{corr}$, along with elevated $E_\mathrm{corr}$ and $R_\mathrm{p}$ values, indicating effective surface passivation while maintaining a compact and defect-free surface.

\par Although as-deposited coatings with optimized composition achieved ICR values below $10~\mathrm{m}\Omega\text{·}\mathrm{cm}^2$, satisfying the U.S.~DOE criterion, the potentiostatic polarization at elevated potentials resulted in a significant increase in ICR to several hundred $\mathrm{m}\Omega\text{·}\mathrm{cm}^2$. However, applying just a 5~nm Pt overlayer to the Ti--Nb--O coatings achieved ICR values of approximately $5~\mathrm{m}\Omega\text{·}\mathrm{cm}^2$ after the potentiostatic polarization~--- a reduction in Pt thickness by one to two orders of magnitude compared to conventional approaches.

\par Our findings highlight the potential of these coatings as a durable, cost-effective, and sustainable solution for PEM electrolyzers with significantly reduced reliance on precious metals. While the present work demonstrates performance on flat stainless steel sheets, the coating strategy was developed with broader applicability in mind, motivating future research on adapting and validating this approach for more complex BPP and PTL geometries in operating electrolyzers. For steel PTLs, additional studies should focus on testing whether preliminary treatment to fully cover all internal surfaces with a protective layer is required.

\section*{Acknowledgments}
This work was supported by the project Quantum materials for applications in sustainable technologies (QM4ST), funded as project No. CZ.02.01.01/00/22\_008/0004572 by Programme Johannes Amos Comenius, call Excellent Research.


\printcredits

\bibliographystyle{elsarticle-num-names}

\bibliography{Refs-DavidKolenaty}


\includepdf[pages=-]{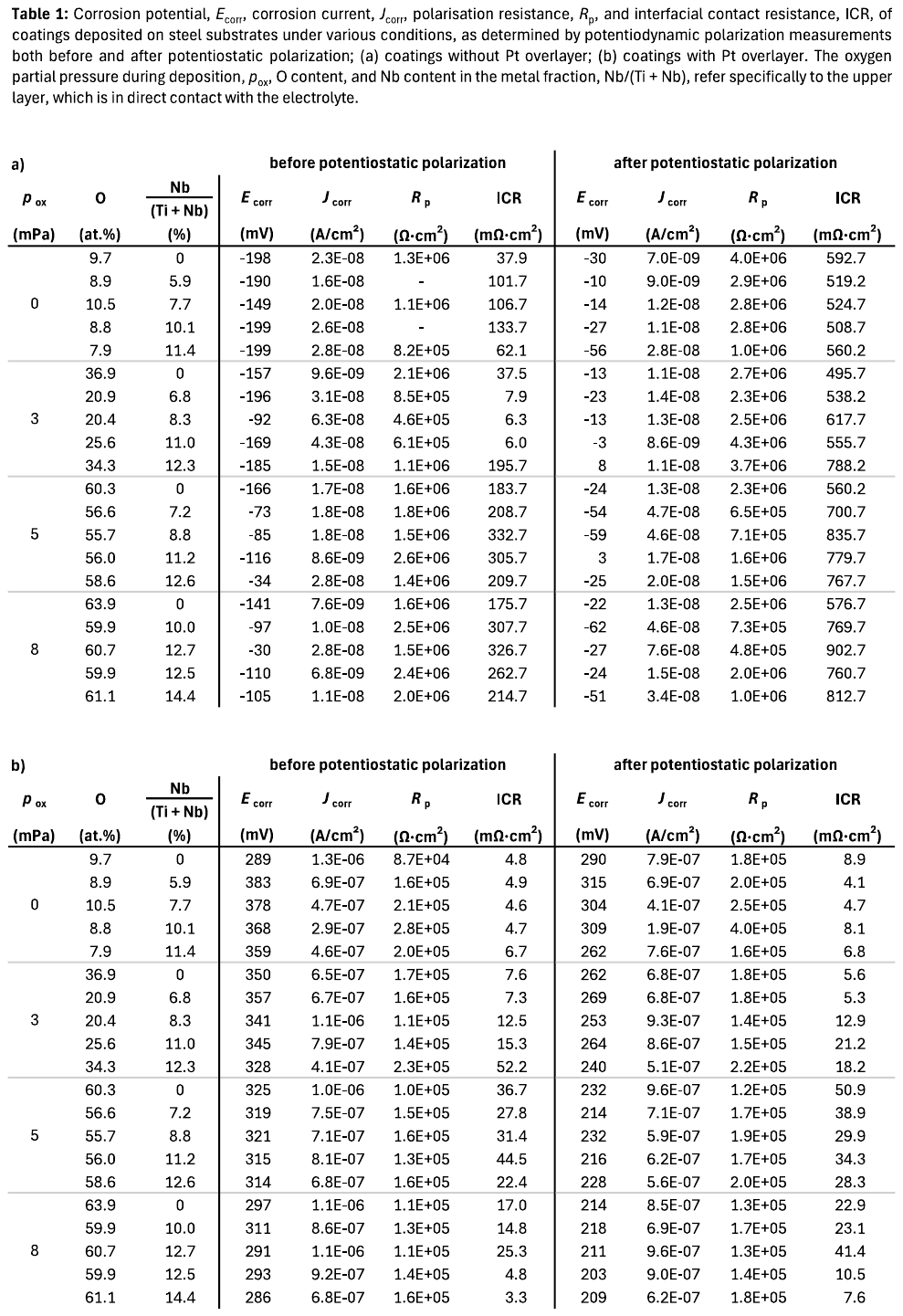}

\end{document}